\documentclass[twocolumn,pra,aps,showpacs,superscriptaddress]{revtex4}
\usepackage{amsmath,amssymb,amsfonts,bbm,graphicx}

\newcommand{\ket}[1]{| #1 \rangle}

\newcommand{\ketbra}[2]{  | #1 \rangle \langle #2 |}

\renewcommand{\[}{\begin{equation}}
\renewcommand{\]}{\end{equation}}

\begin{document}

\title{Tunable Entanglement, Antibunching and Saturation effects in Dipole Blockade}

\author{J. Gillet}
\affiliation{Institut de Physique Nucl\'eaire, Atomique et de Spectroscopie, Universit\'e de Li\`ege, 4000 Li\`ege, Belgium}
\author{G. S. Agarwal}
\affiliation{Department of Physics, Oklahoma State University, Stillwater, OK 74078-3072, USA}
\author{T. Bastin}
\affiliation{Institut de Physique Nucl\'eaire, Atomique et de Spectroscopie, Universit\'e de Li\`ege, 4000 Li\`ege, Belgium}

\date{\today}

\begin{abstract}
We report a model that makes it possible to analyze quantitatively the dipole blockade effect on the dynamical evolution of a two two-level atom system driven by an external laser field. The multiple excitations of the atomic sample are taken into account. We find very large concurrence in the dipole blockade regime. We further find that entanglement can be tuned by changing the intensity of the exciting laser. We also report a way to lift the dipole blockade paving the way to manipulate in a controllable way the blockade effects. We finally report how a continuous monitoring of the dipole blockade would be possible using photon-photon correlations of the scattered light in a regime where the spontaneous emission would dominate dissipation in the sample.
\end{abstract}


\maketitle
Dipole-dipole interactions between atoms or molecules affect profoundly the light absorption that occurs in matter~\cite{Var92}. They have been known for several years to give rise to fascinating applications in quantum information science like quantum logic operations in neutral atoms~\cite{Jak00, Pro02} or entanglement production in mesoscopic ensembles~\cite{Luk01, Hett02, Saf02}. The level shifts associated with those interactions can strongly modify the laser excitation of adjacent atoms, up to a complete suppression of more than one excitation in nearby atoms. In this so-called \emph{dipole blockade} effect, the first excited atom prevents any further excitation in a confined volume by shifting the resonance for its non-excited neighbors, resulting in production of singly excited collective states~\cite{Luk01}. In the past years, evidence for the dipole blockade effect has been obtained with samples of Rydberg atoms because of their strong long-range interaction~\cite{Tong04, Singer04, Cubel05, Vogt06}. An analogous photon blockade effect in an optical cavity has also been reported~\cite{Bir05}. Recently Rabi oscillations between the ground state of a pair of Rydberg atoms and the single excited symmetric collective state has been observed for atoms located more than a few micrometers away~\cite{Urb09, Gae09}. In all those fascinating achievements, the residual effects resulting from possible multiple excitations of the atomic sample are usually not discussed although they cannot be eliminated totally. This motivates a deeper quantitative analysis of the dipole blockade phenomenon to optimize its occurrence and understand its possible limitations~\cite{Vogt06, Poh09}. In the present paper, we report a model aiming at yielding quantitative results as function of the most important experimental parameters including the dipole-dipole interaction strength. The system investigated is a two two-level atom system continuously driven by an external laser field. We report several characteristics of dipole blockade including a tunable steady state entanglement production and a saturation effect in strong driving condition. We also report how a continuous monitoring of the dipole blockade could be obtained with help of the photon-photon correlation signal of the scattered light in a regime where the spontaneous emission would dominate the dissipation effects of the sample.

We consider two atoms at fixed positions $\mathbf x_1$ and $\mathbf x_2$ with internal levels $|e\rangle$ and $|g\rangle$, dipolar transition frequency $\omega = 2 \pi c / \lambda$, and single atom spontaneous emission rate $2 \gamma_s $. The system is conveniently described in the Dicke basis $|ee\rangle$, $|gg\rangle$, $|s\rangle \equiv (|eg\rangle + |ge\rangle)/\sqrt{2}$ and $|a\rangle \equiv (|eg\rangle - |ge\rangle)/\sqrt{2}$. We consider that the two atoms strongly interact when in state $|ee\rangle$ resulting in a shift $\hbar \delta$ of this doubly excited state. They are driven by a resonant external laser field with wave vector $\mathbf k_L$ and Rabi frequency $2\Omega$. In the rotating-wave approximation, the coherent evolution of the system is described by the interaction Hamiltonian
\begin{equation}
H= \hbar \delta \ketbra{ee}{ee} + \hbar \Omega \left(e^{i \mathbf k_L \cdot \mathbf x_1} S^+_1 + e^{i \mathbf k_L \cdot \mathbf x_2} S^+_2 + \mbox{h.c.} \right),
\end{equation}
where $S_i^+ = (S_i^-)^{\dagger}$ ($i = 1,2$) is the atom raising operator $|e\rangle_i \langle g|$ and the term $\hbar \delta \ketbra{ee}{ee}$ accounts for the shift of the doubly excited state of the system induced by the dipole-dipole interaction. Throughout this paper $\mathbf{k}_L$ is supposed to be perpendicular to the two-atom line and the reference frame is properly chosen so as $\mathbf k_L~\cdot~\mathbf x_1 = \mathbf k_L~\cdot~ \mathbf x_2 = 0$. When considering dissipation in the Markov and Born approximation, the time evolution of the system is governed by the master equation
\begin{equation}
\label{eq-ME}
\dot\rho = - \frac{i}{\hbar} [H,\rho] - \gamma \sum_{i=1}^2 (S^+_i S^-_i \rho +\rho S^+_i S^-_i -2 S^-_i \rho S^+_i),
\end{equation}
where $\gamma=\gamma_s+\gamma_d$ with $2\gamma_d$ the dissipation rate modeling non-radiative dissipative effects in the sample. We consider that the two atoms are separated by more than the transition wavelength $\lambda$ so that we can neglect the imbalance among the decay rates of the Dicke states $|s\rangle$ and $|a\rangle$~\cite{Jvz06}. This situation is encountered in most recent experiments, like in Ref.~\cite{Urb09} where the atoms are located more than $20\lambda$ away.

\begin{figure}
\center
\includegraphics[width=4.5cm]{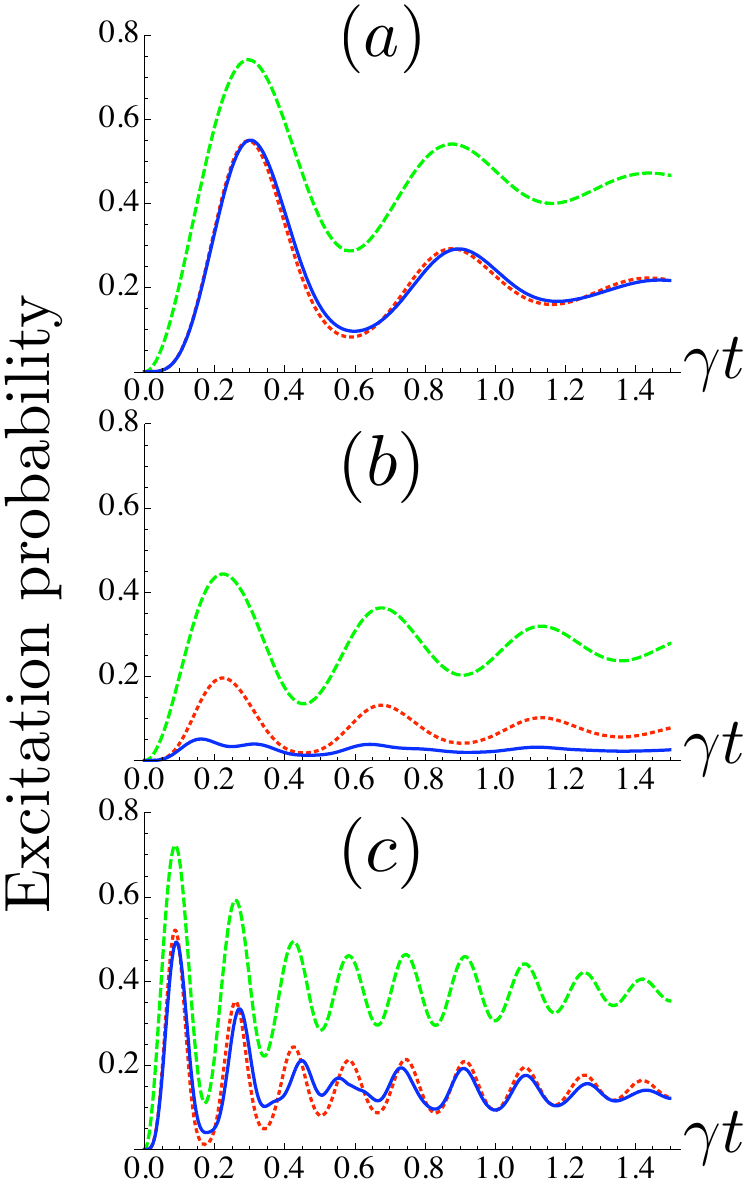}
\caption{(Color online) Time evolution of the excitation probability $P_e$ (dashed green curve), its square (dotted red curve) and the probability $P_{ee}$ of having both atoms excited (blue curve) (a~: $\Omega/\gamma = 5$, $\delta/\gamma = 5$; b~: $\Omega/\gamma = 5$, $\delta/\gamma = 30$; a~: $\Omega/\gamma = 15$, $\delta/\gamma = 30$). The dipole blockade effect is well marked in case b where $P_{ee} \ll P_e^2$.}
\label{figExcitationProbability}
\end{figure}

In presence of the dipole blockade mechanism, the doubly excited state $|ee\rangle$ is expected to be poorly populated though not totally depopulated. This is illustrated quantitatively in Fig.~\ref{figExcitationProbability} where we compare the time evolution of the square of the probability $P_e = \langle e | \mathrm{Tr}_1 \rho | e \rangle = \langle e | \mathrm{Tr}_2 \rho | e \rangle$ of having one of the two atoms excited with the probability $P_{ee} = \langle ee | \rho | ee \rangle$ of finding both atoms excited, considering them initially in the ground state. When the dipole-dipole interaction is not strong enough (case a of Fig.~\ref{figExcitationProbability}), it has negligible effect and the atoms react as independent systems~: $P_{ee} \simeq P_e^2$. For greater dipole-dipole interaction (case b), the double excitation is blocked and the population of the $|ee\rangle$ state remains at insignificant levels, though not zero. More importantly the double excitation probability $P_{ee}$ is much lower than $P_e^2$, giving a direct signature of the blockade mechanism. When the laser intensity is increased (case c), we observe that $P_{ee}$ is again very similar to $P_e^2$. The population blockade is lifted and the atoms behave again as if they were independent without mutual influence. The dipole blockade effect can thus be circumvented by using strong laser fields. Case b exhibits a similar behavior of the system as that observed experimentally in Ref.~\cite{Gae09}.

\begin{figure}
\centering
\includegraphics[width=4.5cm]{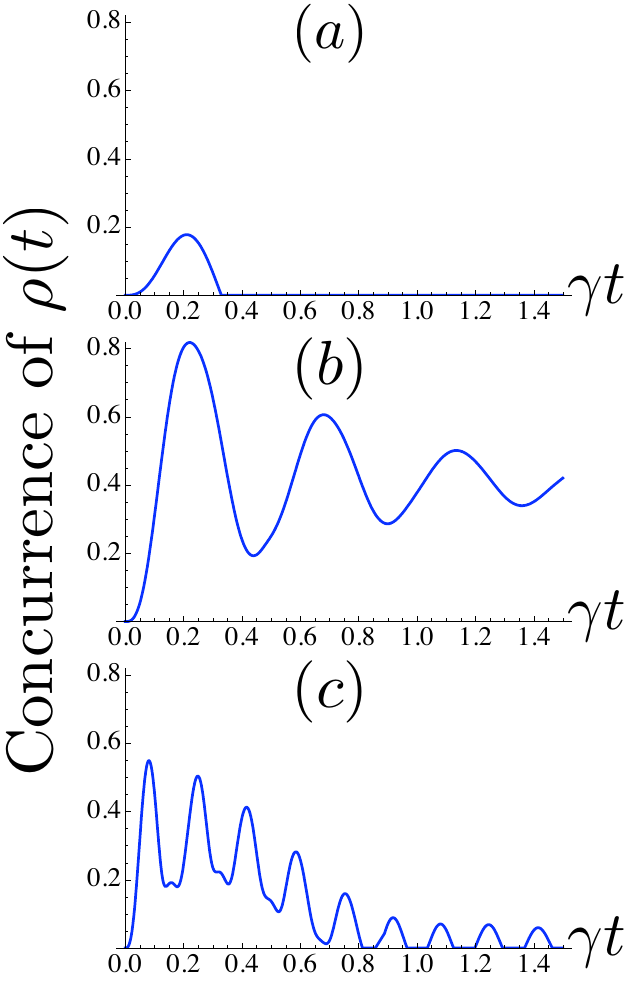}
\caption{(Color online) Time evolution of the concurrence $C$ of the two-atom system (a~: $\Omega/\gamma = 5$, $\delta/\gamma = 5$; b~: $\Omega/\gamma = 5$, $\delta/\gamma = 30$; a~: $\Omega/\gamma = 15$, $\delta/\gamma = 30$).}
\label{figConcurrence}
\end{figure}

The experimental results reported in Refs.~\cite{Urb09, Gae09} clearly imply the entanglement in the two atom system. We can quantify such an entanglement. From the master equation we can obtain the complete time dependent density matrix which then can be used to compute the well known measure of entanglement~: the concurrence~\cite{Woo98}. We show the results in Fig~\ref{figConcurrence}. The concurrence is maximized when the dipole blockade mechanism is itself optimized. In case a, the dipole-dipole interaction is too weak and the two-atom system behaves as a collection of independent atoms. No significant entanglement is produced. In case b, the dipole blockade prevents the doubly excited state to be significantly populated and the two-atom system shares a collective single excitation. More population in the entangled $(|eg\rangle + |ge\rangle)/\sqrt{2}$ state is expected and significant amounts of entanglement are produced. In case c, the dipole blockade is lifted and more population in the separable doubly excited state is expected. The concurrence is again less important than in case b.

The two-atom state $\rho$ subjected to the master equation~(\ref{eq-ME}) always stabilizes after a finite time around a steady state that we denote $\rho^{SS}$. The steady state is found by equating the right-hand term of Eq.~(\ref{eq-ME}) to zero. We get in the Dicke basis $\{|ee\rangle, |s\rangle, |a\rangle, |gg\rangle\}$
\begin{widetext}
\begin{equation}
\rho^{SS} =  \frac{1}{16 \Omega^4 + (4 \Omega^2+\gamma^2) | \alpha |^2}
\left( \begin{array}{cccc}
4\Omega^4 & 2 \sqrt{2} \Omega^3 \alpha & 0 &  -2 i \Omega^2 \gamma \alpha \\
2 \sqrt{2} \Omega^3 \alpha^*& 2 \Omega^2 (2 \Omega^2+|\alpha|^2) &  0    & \sqrt 2 \Omega (2 \Omega ^2 \alpha-i \gamma |\alpha|^2 ) \\
0 & 0   &   4\Omega^4 & 0  \\
2 i \Omega^2 \gamma \alpha^* & \sqrt 2 \Omega (2 \Omega ^2 \alpha^* +i \gamma |\alpha|^2 )  & 0 & 4 \Omega^4 + (2 \Omega^2+\gamma^2) | \alpha |^2
\end{array} \right),
\end{equation}
\end{widetext}
where $\alpha=-(\delta+2i \gamma)$.

\begin{figure}
\centering
\includegraphics[width=180 pt]{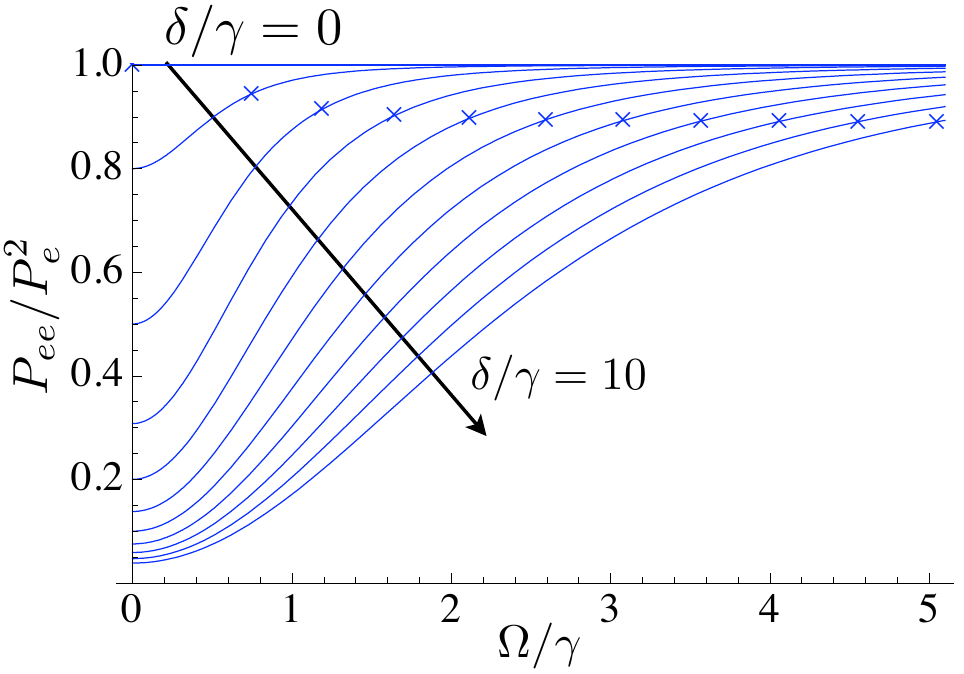}
\caption{(Color online) Plots of $P_{ee}/P_e^2$ with respect to $\Omega/\gamma$ for all integer values of $\delta/\gamma$ from 0 to 10. The crosses indicate for each curve the values of $\Omega/\gamma$ above which the steady state is separable.}
\label{figPeePe2SS}
\end{figure}

In the steady state regime, the population of the doubly excited state $|ee\rangle$ decreases when $\delta$ increases. This is the usual dipole blockade effect where one excited atom prevents the excitation of a nearby atom. This effect is counterbalanced by an increase in the laser intensity. The dipole blockade effect is lifted with use of higher laser intensity. The ratio between the steady state double excitation probability $P_{ee}$
and the square of the single excitation probability $P_e$ reads
\begin{equation}
\label{ratioPeePe2}
\left.\frac{P_{ee}}{P_e^2}\right|_{SS} =  \frac{64 \Omega^4 + 4 (4 \Omega^2+\gamma^2) | \alpha |^2}{ (8 \Omega^2+|\alpha|^2)^2 }.
\end{equation}
In absence of the dipole-dipole interaction ($\delta = 0$) this ratio is trivially equal to 1. This is obviously expected from the absence of correlation in the two-atom system in this case. When increasing $|\delta|$ the ratio monotonically decreases. This is a clear signature of the increasing correlation induced by the stronger and stronger dipole-dipole interaction shifting more and more the doubly excited state. We show more quantitatively the behavior of this ratio for different values of $\delta/\gamma$ with respect to the field intensity in Fig.~\ref{figPeePe2SS}. It is quite clear that for weak intensities of the
field, the dipole blockade regime is dominant as there is less and less population in the $\ket{ee}$ state as $\delta/\gamma$ increases.
However, increasing the field intensity has the effect of repopulating the $\ket{ee}$ state and therefore lifting the dipole blockade.

The concurrence of the steady state reads
\begin{equation}
C(\rho^{SS}) = \mbox{Max}\left\{0, \frac{\sqrt 2 \Omega^2 (\lambda_+ - \lambda_-) -
8 \Omega^4}{16 \Omega^4 + (4 \Omega^2+\gamma^2) | \alpha |^2} \right\},
\end{equation}
with
\begin{equation}
\lambda_{\pm} = \sqrt{8 \Omega^4 + \delta^2 |\alpha|^2 \pm \delta |\alpha| \sqrt{16 \Omega^4 + \delta^2 |\alpha|^2}}.
\end{equation}

\begin{figure}
\centering
\includegraphics[width=180 pt]{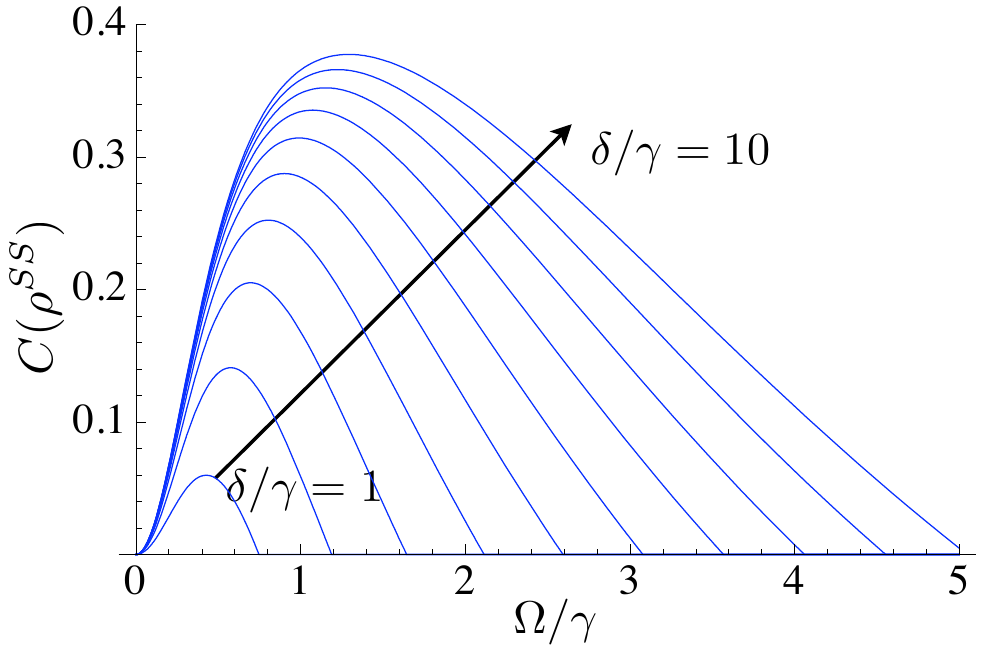}
\caption{(Color online) Plots of $C(\rho^{SS})$ with respect to $\Omega/\gamma$ for integer values of $\delta/\gamma$ from 1 to 10.}
\label{figConcSS}
\end{figure}

In absence of dipole-dipole interaction ($\delta = 0$), the steady state is not entangled. No entanglement is produced in this configuration since the two atoms behave as independent systems. This highlights the fundamental role of the dipole blockade mechanism for long-term entanglement production of the two-atom system. For increasing values of $\delta$, we show in Fig.~\ref{figConcSS} the concurrence of the steady state with respect to the field intensity. The amount of
long-term entanglement in the system is clearly tunable with the laser intensity and can be reasonably high for well adjusted
values of $\delta$ and $\Omega$. When the intensity of
the field increases and lifts the dipole blockade, the amount of
entanglement decreases accordingly. The steady
state is entangled as long as
\begin{equation}
\label{eqMaxOmega}
0 < 4 \Omega^2 < \delta|\alpha|. 
\end{equation}
That upper limit on $\Omega$ is pointed on each plot of Fig.~\ref{figPeePe2SS}.

The photon-photon correlation signal gives information that is not contained in intensity measurements and is a good probe for the quantum nature of the investigated processes. In our setup, the photon-photon correlation function is given by~\cite{Sko01, Jvz06}
\begin{equation}
\label{g2probab}
   g^{(2)}({\mathbf r}_1,t;{\mathbf r}_2, t + \tau) = \frac{P({\mathbf r}_2, t + \tau|{\mathbf r}_1,t)}{P({\mathbf r}_2,t)},
\end{equation}
where $P({\mathbf r}, t)$ is the probability of detecting a photon
at position ${\mathbf r}$ and time $t$, and $P({\mathbf r}_2, t +
\tau|{\mathbf r}_1,t)$ the conditional probability of finding a
photon at ${\mathbf r}_2$ and $t + \tau$ assuming that a photon at
${\mathbf r}_1$ and $t$ has been recorded. The probabilities $P({\mathbf r}_1,t)$ and $P({\mathbf r}_2, t +
\tau|{\mathbf r}_1,t)$ are given by $\left\langle D^{\dagger}({\mathbf r}_1) D({\mathbf r}_1
) \right\rangle_{\rho(t)}$ and
$\left\langle D^{\dagger}({\mathbf r}_2) D({\mathbf r}_2) \right\rangle_{\rho'(t + \tau;{\mathbf r}_1,t)}$, respectively,
where $\rho(t)$ is the density operator of the two-atom system at
time $t$, $\rho'(t + \tau;{\mathbf r}_1,t)$ is the density
operator at time $t + \tau$ assuming a photon has been detected at
point ${\mathbf r}_1$ and time $t$, and $D({\mathbf r})$ is the
photon detector operator $S_1^{-} + e^{i \phi({\mathbf r})}
S_2^{-}$, where $\phi ({\mathbf r}) = k_L \hat{{\mathbf r}} \cdot ({\mathbf x}_1 - {\mathbf x}_2)$ and $\hat{{\mathbf r}} = {\mathbf r}/r$.

\begin{figure}
\centering
\includegraphics[width=5 cm]{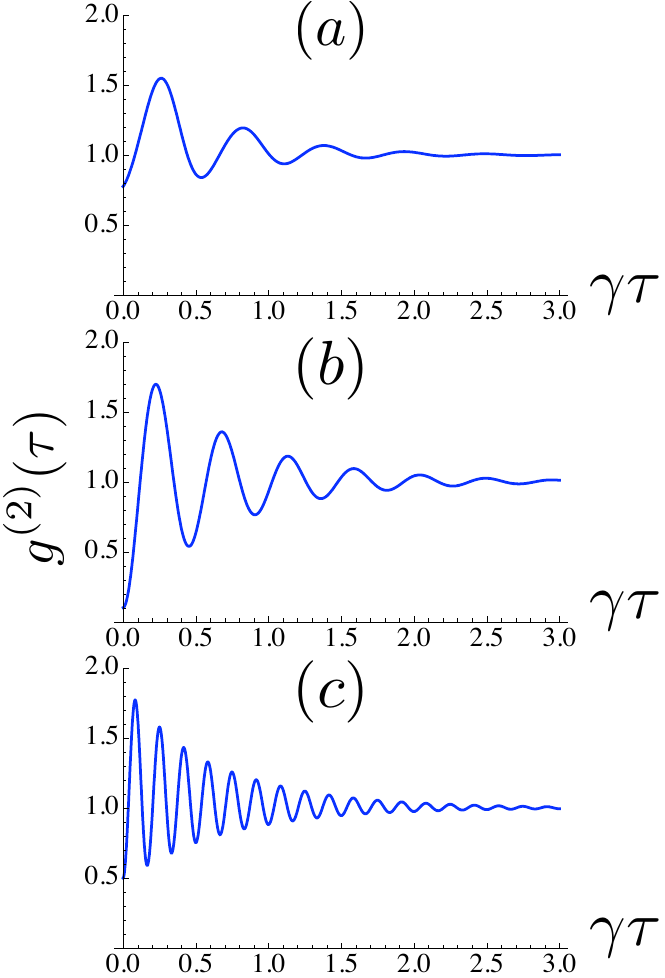}
\caption{(Color online) Second order correlation function $g^{(2)}(\tau)$. (a~: $\Omega/\gamma = 5$, $\delta/\gamma = 5$; b~: $\Omega/\gamma = 5$, $\delta/\gamma = 30$; c~: $\Omega/\gamma = 15$, $\delta/\gamma = 30$).} \label{figg2}
\end{figure}

We show in Fig.~\ref{figg2} the photon-photon correlation function (\ref{g2probab}) with respect to $\tau$ in a time $t$ when the system is in the steady state and where the two detectors are located such that $\phi(\mathbf r_1) = \phi(\mathbf r_2) = 2 n \pi$ with $n$ an integer number. Although this is not yet the case in the first experimental observations of the dipole blockade manifestations~\cite{Urb09, Gae09}, we consider here a regime where the spontaneous emission dominates all dissipative effects in the atomic sample ($\gamma \approx \gamma_s$). Similar experimental parameters to those used in Figs.~\ref{figExcitationProbability} and \ref{figConcurrence} have been considered. For low dipole-dipole interaction (case a), a usual antibunching behavior of the scattered photons is observed~\cite{Sko01}. For higher dipole-dipole interaction (case b), the antibunching of the scattered photons is much more marked as the value of the correlation function for $\tau = 0$ is much smaller with a much higher slope with respect to $\tau$. The dipole blockade enhances the antibunching behavior. For higher laser intensities (case c), $g^{(2)}(\tau = 0)$ increases again and the dipole blockade effect is less marked.

For $\tau = 0$ and considering the time $t = 0$ when the system is in the steady state, we get
\begin{align}
\label{g2total}
g^{(2)}(\mathbf{r}_1, 0 ; \mathbf{r}_2,0) = & \\
& \textrm{~\hspace{-2.5cm}} \frac{4(16\Omega^4+(4 \Omega^2+\gamma^2) |\alpha|^2) \cos^2((\phi_1-\phi_2)/2)}{(8\Omega^2+|\alpha|^2(1 + \cos{\phi_1}))(8\Omega^2+|\alpha|^2(1 + \cos{\phi_2}))}, \nonumber
\end{align}
with $\phi_i \equiv \phi(\mathbf r_i)$ ($i = 1,2$). Some particular detector positions are worth investigating. When $\phi_1 = \phi_2 = (2n+1)\pi$ with $n$ an integer, the photon-photon correlation function (\ref{g2total}) exhibits a simple dependence to the dipole blockade parameter $\delta$, that appears only in the numerator through a quadratic dependence. The most interesting regime is reached when $\phi_1 = \phi_2 = (2n+1)\pi/2$. In this case,
\begin{equation}
g^{(2)}(\mathbf{r}_1, 0 ; \mathbf{r}_2,0) = \left.\frac{P_{ee}}{P_e^2}\right|_{SS}
\end{equation}
and the photon-photon correlation function identifies to the ratio (\ref{ratioPeePe2}) between the steady state double excitation probability and the square of the single excitation probability. This ratio is a direct measure of the dipole blockade effect. The more it diverges from 1, the more intense the dipole-dipole interactions are. For those particular detector positions, the coincident photon-photon correlation signal monitors quantitatively the dipole blockade in the two-atom sample. This monitoring works continuously as long as the system is permanently driven in its steady state and scatters the laser light.

As a conclusion, we have provided a model able to analyze quantitatively the dipole blockade effect on the dynamical evolution of a two two-level atom system. We have shown that the dipole blockade is an efficient mechanism for production of significant long-term entanglement in the steady state of the system when it is continuously driven by a resonant laser field. This long-term entanglement non-existent in absence of dipole blockade is tunable with the laser intensity. We have proven that the effect of the dipole blockade can be lifted in strong driving conditions.
Finally we have shown that for particular detector positions, the photon-photon correlation function could continuously monitor the dipole-dipole interaction between the two atoms in a regime where the spontaneous emission would dominate all dissipative effects in the atomic sample. That would provide an efficient tool in the analysis of the occurrence of the dipole blockade.

\acknowledgments{J.~G. thanks the Belgian F.R.S.-FNRS for financial support.}

\end{document}